# "Acoustical hooks: a new subwavelength self-bending beam"

Constanza Rubio[(1)*], Daniel Tarrazó-Serrano[(1)], Oleg V. Minin[(2), (3)], Antonio Uris[(1)] and Igor V. Minin[(2), (3)]

(1) Centro de Tecnologías Físicas: Acústica, Materiales y Astrofísica, Universitat Politècnica de València, Camino de Vera s/n, 46022 Valencia, Spain

(2) Tomsk Polytechnic University, 36 Lenin Avenue, Tomsk, 634050, Russia

(3) Tomsk State University, 30 Lenin Avenue, Tomsk, 634050, Russia

* Corresponding author: E-mail: crubiom@fis.upv.es

In this work, we report the observation of a new type of near-field curved acoustic beam different from the Airy-family beams both through simulations and experiment. This new self-bending acoustical beam is generate from a rectangular trapezoid of a polymer material immersed in water and has unique features. The radius of curvature of acoustical hook is less than the wavelength, it is represents the smallest radius of curvature ever recorded for any acoustical beams. The origin of this curved beam is in the vortices of intensity flow that appear inside the solid due to the conversion of the incident longitudinal wave mode to a shear wave in a solid. These vortices redirect the intensity flow, which causes the beam to bend. These results may be potentially useful when an object, that is in the path of the beam, must be avoid, such as in cancer treatments in which tumors are found behind bones such as ribs. It could also have potential applications in particle manipulations.

**Introduction.**

It is well known since antiquity that waves propagate in a straight line. However, the possibility of a wave propagating along a curved path was suggested and observed experimentally in optics in 2007 [1, 2]. The concept arises from quantum mechanics in which, in the absence of an external force, a wave packet can be accelerated as long as the quantum wave function follows an Airy function profile [3]. The exact solution of the Schrödinger equation and the paraxial optical wave equation gives rise to a beam that accelerates in a certain trajectory, that is to say it bends, and being free diffraction. This type of beam was called an Airy beam. It could be noted that the main lobe of a finite energy Airy beam is not observable directly behind the cubic phase element and a transition region exists, where the initial intensity distribution of the incoming beam is transformed into the distinct Airy pattern [4]. Since then, Airy beams in optics have attracted great attention from researchers and possible applications have been proposed [5-14].

In the last years, the studies on optical Airy beams have be extended to acoustic waves, which have attracted great attention [15-23]. It could be noted that acoustic waves are elastic and longitudinal in nature waves. However, electromagnetic (optical) waves are transverse due to, both the electric and magnetic fields, are mutually perpendicular to the direction that the wave travels. For the generation of an acoustic Airy-like beams in liquids different methods were investigated: a piston transducer with a corrugated face, specifically tailored [15], active elements controlled by electronics in air [16] and water [17-18], or meta-materials with well-tuned parameters [19-21], which requires the errors of its scale to be much smaller than the wavelength. Some other types of waves which satisfy the paraxial wave equation, such as water waves [22,23] generated by phase modulating a single projector using a tailored acoustic phase mask and acoustic Airy beams by using a zero-index medium to provide the required phase profile [24] were also considered, among others. An acoustic wave propagating in a curved path could have many applications in ultrasonic imaging, medical ultrasound and particle manipulation [16, 23, 25]. At present, the designs that have been proposed are relatively complex, which limits their application. Moreover, all of well known acoustic self-bending beams has a curvature more than at least of several wavelength.

On the other hand, recently, a new family of near-field localized curved optical beams has been discovered by Minin and Minin [26], which differs from the family of Airy beams. This new family is able to form in the near-field zone the so-called "photonic hook", which is self-bending (transversely accelerate) throughout propagation in near field in free space [27]. Unlike the Airy-like beam, the photonic hook is created by focusing a plane wave through a dielectric particle combination of a wedge prism and a cuboid, and because of this shape with broken symmetry, the time of the full oscillation phase of an optical wave varies in a particle unevenly. As a result, a curved light beam is produced at the exit from the particle (near it shadow surface). Photonic hook are unique in that their radius of curvature is substantially smaller than the wavelength [27, 28]. That is, such structured light beams have the maximum acceleration among the known curvilinear beams - the curvature of electromagnetic waves with such a small radius were described for the first time in Refs. [26,27] and confirmed experimentally in this year 2019, by Minin and Minin and colleagues in THz [28]. The phenomenon of photonic hook was demonstrated also for surface plasmon wave [29] realized in the in-plane of the interface despite the strong energy dissipation at metal surface. These results have generated expectation in the scientific community [30].

Not long ago, and by using the formal analogy between electromagnetic and acoustic waves, Minin and Minin [31] transferred the idea of photonic jet effect [32] to acoustic waves. They demonstrated the existence of the phenomenon of the phononic jet in the shadow area of a penetrable sphere, which they called "acoustojet" (AJ) [33]. For the experimental demonstration of the AJ phenomenon, a Rexolite® sphere with a diameter of 8 wavelengths immersed in water and at a frequency of 1.01 MHz was used [34]. Rexolite® was chosen so that there would not be a significant difference between the impedance of the material and that of the surrounding medium, since otherwise the intensity of the acoustic jet would be relatively low [35].

Taking into account this formal analogy between electromagnetic and acoustic waves [31] it can be expected that the photonic hook formation method [27] can be adapted to acoustics. However, on the other hand, unlike electromagnetic waves, in the acoustics field, an additional transverse velocity of sound (shear wave) has to consider in solids; this fact causes them to become anisotropic. Therefore, the possibility of implementing the effect of the acoustic hook is not so obvious. So, to this day, not an experiment was ever reported on acoustic hook (AH) effect in liquids. Here, we are going to introduce a simple method to generate the acoustic hook, which can be used to bend, as desired, the acoustic beam in near field without sophisticated control-devices. Its simplicity would make the fabrication process easier and less expensive.

In this work, we show both by simulations and experimentally that a rectangular trapezoidal Rexolite® particle on which a plane ultrasonic wave impinges can generate a curved acoustic beam with unique properties, with radius of curvature substantially smaller than the wavelength in water. In addition, the influence of the inner angle of the rectilinear trapezoid in the generation of the curved beam is analyzed numerically.

**Results**

Figure 1 shows the simulation results obtained to describe the normalized sound pressure $\frac{|P|^2}{|P_i|^2}$ distributions (where p and $p_i$ are the pressure and the incident sound pressure at each point, respectively) in XZ planes for Rexolite® cuboid particle and rectangular trapezoidal particles with different interior angles. As can be seen in Figure 1 (a), an acoustic field enhancement appears on the shadow surface of the cuboid particle. This phenomenon is the AJ. When the cuboid is replaced by a rectangular trapezoid with an interior angle of 10º, it is observed that the shape of the AJ is slightly bent (see Figure 1(b)). As the interior angle increases, taking values of 15º (Figure 1(c)) and 20º (Figure 1(d)), it is observed that the curvature of the AJ increases as well as the normalized sound pressure $\frac{|P|^2}{|P_i|^2}$. For interior angles values higher than 20º, it is observed that the effect weakens (Figure 1(e)-(f)). To quantify how much the AJ is bent, the curvature is defined. As defined in Refs. 24-25, the AJ curvature is determined by a midline $L_{jet}$ aided by an angle β between the two lines, which link the start point with the inflection point and the inflection point with the end point, respectively, see Figure 2. Table I summarizes the AJ curvature as a function of rectangular trapezoid interior angle α. As can be observed, the cuboid particle, that has symmetry (α = 0º), produces a symmetric AJ, that is, without curvature (β = 0º). When the symmetry of the particle is broken using a rectangular trapezoid, the AJ ceases to be symmetrical and begins to bend. As the interior angle α increases, the angle of curvature β decreases, it should be note that for an angle α of 30º the effect is no longer perceptible.

Table I. Curvatures β of AH for different interior angle α

| α | 0º | 10º | 15º | 20º | 25º | 30º |
|---|----|-----|-----|-----|-----|-----|
| β | 0º | 168º | 150º | 142º | 134º | - |

Interestingly, in the case of acoustic hook only main lobe has a curved shape and the family of curved sidelobes, as in Airy beams are absent. The pressure maximum not located on the shadow surface of the particle, as is the case in optics, and shifted to a distance of about 0.17 of wavelength from the surface of the particle. The minimal FWHM of AJ for cube is located at the shadow surface of cube and is about 1.7λ. In the case of a particle with broken symmetry minimal FWHM of AH is 0.717λ at the shadow surface of particle (i.e. has a subwavelength value) and FWHM is about 0.83λ at the point of maximal pressure along AH.

The technique of ultrasonic immersion was used to experimentally verify the self-bending effect of an AJ. A cuboid particle of side length 3λ, and a rectangular trapezoidal particle obtained from the 3λ side cuboid and with an interior angle of 20º were chosen (see Figure 3(a)-(c)). The experimental results for the cuboid particle and the rectangular trapezoidal particle are shown in Figure 4(a) and 4(b) respectively. Figure 4(b) clearly shows the effect of the formation of a curvilinear region of the acoustic field localization behind a dielectric particle with broken symmetry. It could be observed that an experimental curvature β = 144º is obtained. It must be kept in mind that the simulations carried out and shown in Figure 1(a)-(f) have been carried out considering the incidence of a plane wave. However, the emission of the transducer is not really plane wave, although the particles were at a relatively long distance from the transducer. However, the experimental results agree well with the simulated results, which allow validate the results of the simulations presented in this work.

Table II. Parameters of the experimental acoustic hook in different cross-sections

| Cross-section | $X_{max}/\lambda$ | $Z/\lambda$ | $P/P_{max}$ | $FWHM/\lambda$ |
|---|---|---|---|---|
| I | 0.0835 | 0.333 | 0.9153 | 0.707 |
| II | -0.0832 | 0.333 | 0.9004 | 0.825 |
| III | -0.0832 | 0.500 | 0.7894 | 0.825 |
| IV | -0.2499 | 0.500 | 0.7292 | 0.825 |
| V | -0.2499 | 0.667 | 0.7189 | 0.557 |
| VI | -0.4166 | 0.667 | 0.7132 | 0.589 |
| VII | -0.4166 | 0.833 | 0.6907 | 0.707 |
| VIII | -0.4166 | 1.000 | 0.5822 | 1.061 |

Table II shows the parameters of the experimental acoustic hook for the rectangular trapezoidal particle. It is observed that the lateral dimensions (FWHM/λ) of the AH are subwavelength and improves the experimental results obtained in the AJ: for the cross-section V, which corresponds to Z/λ = 0.667, an FWHM/λ = 1.169 is obtained of the cuboid particle, while in the case of the rectangular trapezoidal particle an FWHM/λ = 0.557 is obtained. Thus, the phenomenon of AH could provide advantages over the AJ in certain applications since it improves the spatial resolution.

The AH phenomenon can be explained by using the relative intensity ($\vec{I} = p \cdot \vec{u}$, where p is the acoustic pressure and $\vec{u}$ is the particle velocity) flow diagrams in XZ planes for the particles considered. Figure 5(a) shows the relative flow diagrams in XZ plane for the 3λ side cuboid particle. It is observed that there are regions within the cuboid where the intensity lines produce vortices that redirect the intensity flow to different areas of the cuboid. It is observed that there are regions within the cuboid where the intensity lines produce vortices that redirect the intensity flow to different areas of the cuboid [27,28]. Due to the cuboid's symmetry, the distribution of the vortices is symmetrical, so that the AJ obtained is completely symmetrical, that is, it is not curved. However, when the rectangular trapezoidal particle is considered (see Figure 5(b)), it is observed that the distribution of the vortices that redirect the intensity flow inside the particle is no longer symmetric. This produces the curvature of the AJ. The origin of the vortices is in the conversion of an incident longitudinal wave mode to a shear wave in a solid and then back to a longitudinal wave in the water. The particle velocity fields, $\vec{v}$, can be represented by gradients of the scalar function $\phi$ in a fluid [35]

$$\vec{v} = \vec{\nabla}\phi$$

and the curl of a vector field $\vec{\Psi}$ in a solid

$$\vec{v} = \vec{\nabla}\phi + \vec{\nabla} \times \vec{\Psi}$$

Where $\vec{\nabla}\phi$ represents the longitudinal wave and $\vec{\nabla} \times \vec{\Psi}$ represents the shear wave.

The vortexes in the intensity flow are due to shear waves ($\vec{\nabla} \times \vec{\Psi} \neq 0$). In the case of the cuboid particle, there is symmetry in the position of the vortexes, while in the case of the rectangular trapezoidal particle, there is no such symmetry, which causes the AJ bent.

**Discussion**

In this wok, it was shown that acoustic hook phenomenon is observed near the shadow surface of polymer particle on a scale much smaller than known Airy-family beams.

The intensity flow analysis particles suggests that the origin of the AJ bent phenomenon is the vortices that appear inside the particles due to the conversion of the incident longitudinal wave mode to a shear wave in a solid. When the particle is symmetric these vortices are distributed symmetrically, so that the beam does not curve, whereas when the particle is asymmetric, the distribution of the vortices inside the particle no longer symmetrical causing the AJ bending.

The AH as its optical counterpart has unique features of self-bending with radius of curvature that are substantially smaller than the wavelength and represents the smallest radius of curvature ever recorded for any acoustical beams. The AH have the potential to focus around obstacles that are directly in the beam path similar to optical [36,37]. It is interesting to comment that, due to the analogy between electromagnetic and acoustic waves, the curvatures β obtained experimentally in both cases are similar: β = 144º for THz waves [28] and β= 148º in the acoustic case. The mesoscale dimensions and simplicity of the AH are much more controllable for practical tasks and could enable it to be integrated, for example, into lab-on-a-chip platforms and indicating their large-scale potential applications.

Thus, we demonstrated an effective design for the generation of ultrasonic self-bending (or self-accelerating) AH through a dielectric particle with broken symmetry, immersed in water. The design route was well demonstrated by full-wave simulations and validated preliminarily by experiments. To our opinion, such an elegant examples of structured acoustical beams make possible a deeper understanding of wave propagation, and will fuel anticipation and excitement for the next generation of imaging and near-field manipulation.

Therefore, it has been shown that the concept of photonic hook has gone beyond optics [26, 27] and plasmonics [29] and now penetrated acoustics.

**Methods**

**Simulation**

The results of the simulation were obtained using the commercial software COMSOL Multiphysics Modeling © and with 3D modeling. Two resolution modules have been used with the perfectly coupled multiphysics solution system to be able to consider the two propagation speeds (longitudinal and shear speed) in the rectangular trapezoidal particle of Rexolite®. The type of mesh was adjusted to free tetrahedral, and to avoid numerical dispersion, the maximum size of the element was $\lambda/8$.

The host medium was water with typical sound speed (c) and density (ρ) values ($c_{water}$ = 1500 m·s$^{-1}$ and $\rho_{water}$ = 1000 kg·m$^{-3}$). The working frequency selected for the simulations was 250 kHz. The values used to model Rexolite® were longitudinal sound speed $c_{lRexolite}$ = 2337 m·s$^{-1}$, shear sound speed $c_{sRexolite}$ = 1157 m·s$^{-1}$, and density $\rho_{Rexolite}$ = 1049 kg·m$^{-3}$. [28-30].

**Experimental set-up**

The experimental measurements were carried out by using the technique of ultrasonic immersion transmission. A precision automated measurement system was used. This system consists of a fixed-piston ultrasonic transducer used as an emitter (Imasonic, Les Savourots, France) with a central frequency of 250 kHz and an active diameter of 0.032 m and, as a receiver, a polyvinylidene fluoride needle hydrophone (PVDF) (model HPM1 / 1, acoustic precision Ltd., Dorchester, United Kingdom) with a diameter of 1.5 mm ($\lambda/6$, where $\lambda$ is the incident wavelength in water) and a bandwidth of ± 4 dB spanning from 200 kHz to 15 MHz. To post-amplify and digitize the signal was used a digital oscilloscope for PC (model 3224 of Picoscope, Pico Technology, St. Neots, United Kingdom). The scanning was carried out with steps of 1 mm and during the measurements; the water temperature was 18ºC. The experimental set-up are shown in Figures 3(a)-(b).

To demonstrate experimentally the phenomenon of AH generation, a cuboid and a rectangular trapezoidal particle were machined from a cylinder using a numerical control milling machine. Rexolite® was chosen as material to manufacture the samples. Rexolite® is a plastic cross-linked polystyrene that has low difference in acoustic impedance with respect to water. The cuboid has equal side length, $\ell$. The value of side length chosen was $3\lambda$ ($\lambda$ = 6 mm). The rectangular trapezoidal sample is obtained from the $3\lambda$ side cuboid and had an interior angle of 20º, as shown in Figure 3(c).

**Acknowledgements**

This work has been supported by Spanish Ministry of Science, Innovation and Universities (grant No. RTI2018-100792-B-I00). The research was partially supported by Tomsk Polytechnic University Competitiveness Enhancement Program.

**Contributions**

I.V.M. and O.V.M. initiated and supervised this work. C.R. D.T.-S. and A.U. coordinated the experimental measurements and carried out the simulation. All authors have equally contributed to write the manuscript and discussed the results.

**Competing Interests**

The authors declare no competing interests.

**FIGURE CAPTIONS**

Figure 1. Normalized sound pressure normalized sound pressure $\frac{|p|^2}{|p_i|^2}$ distributions in XZ planes for different Rexolite® particles : (a) 3λ side cuboid, and a rectangular trapezoidal particle obtained from the 3λ side cuboid and with an interior angle (b) 10º, (c) 15º, (d) 20º, (e) 25º and (f) 30º.

Figure 2. Scheme of the curved beam formation

Figure 3. Detail of (a) experimental set-up, (b) coordinate axes. (a) Rexolite® particles considered.

Figure 4. Measured sound pressure distribution in XZ planes for Rexolite® particles (a) 3λ side cuboid and (b) rectangular trapezoidal particle obtained from the 3λ side cuboid and with an interior angle of 20º.

Figure 5. Numerical normalized relative intensity flow in XZ planes for (a) 3λ side cuboid and (b) rectangular trapezoidal particle obtained from the 3λ side cuboid and with an interior angle of 20º.

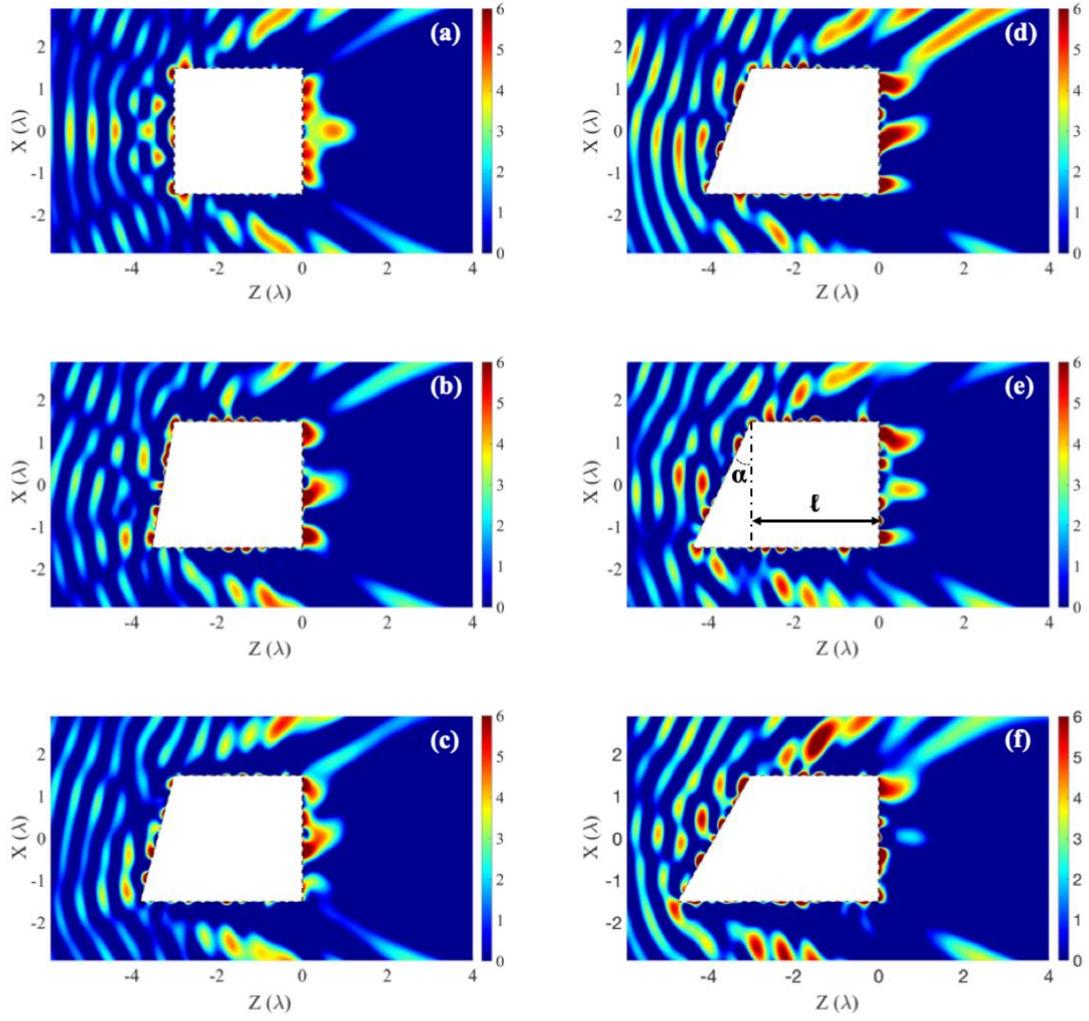

**FIGURE 1**

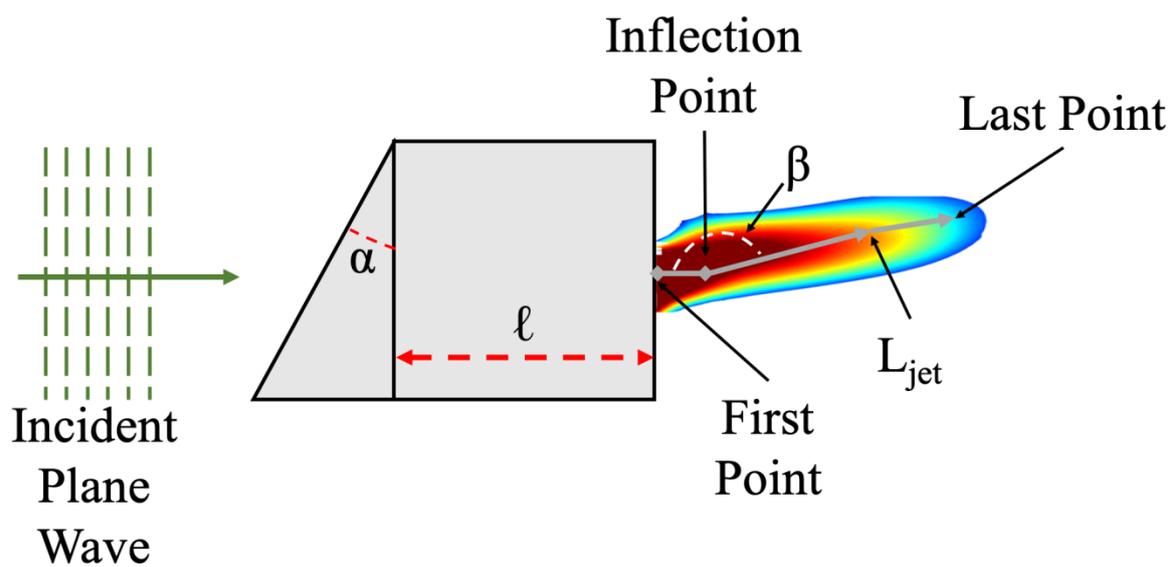

**FIGURE 2**

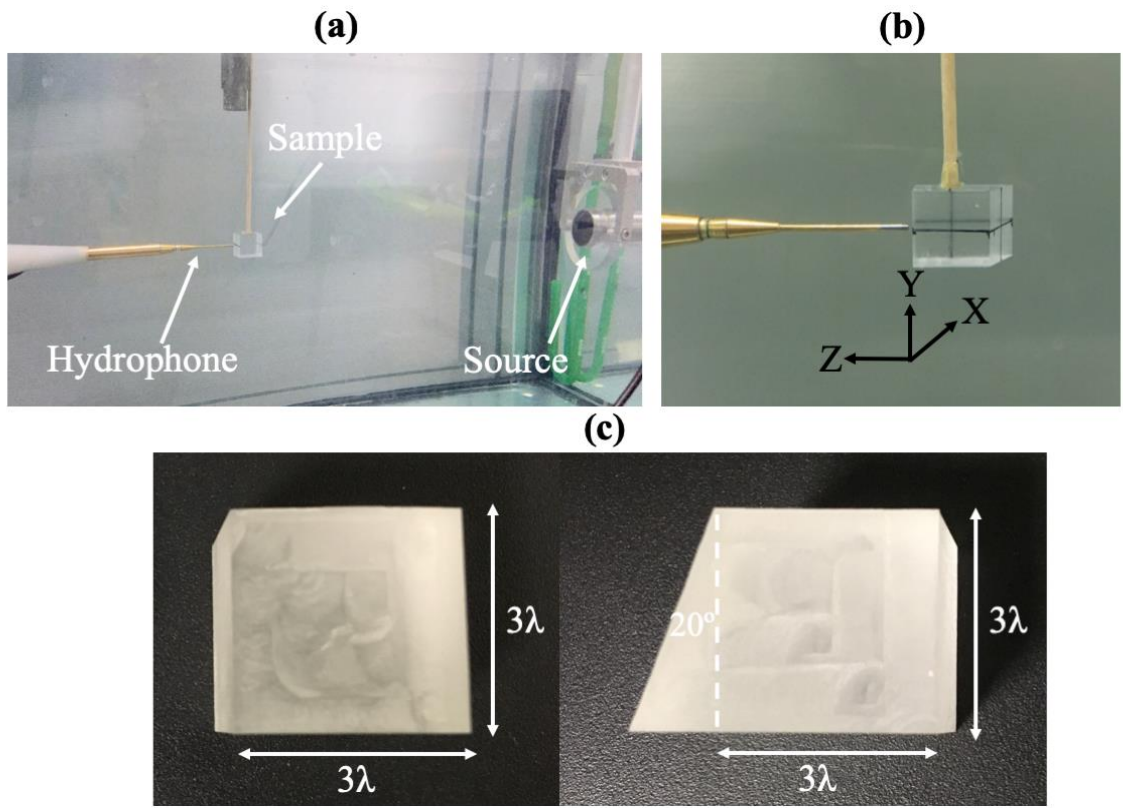

**FIGURE 3**

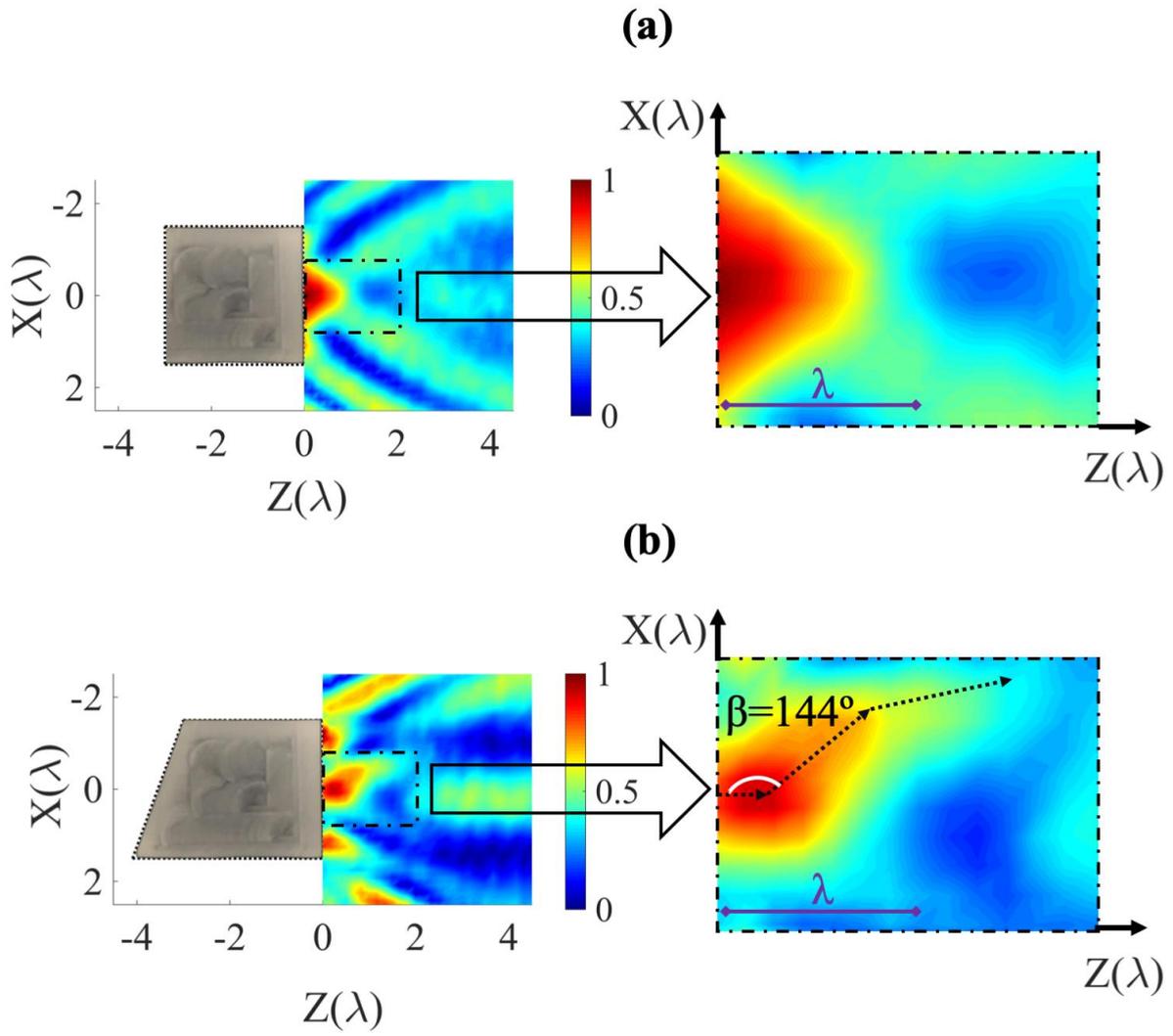

**FIGURE 4**

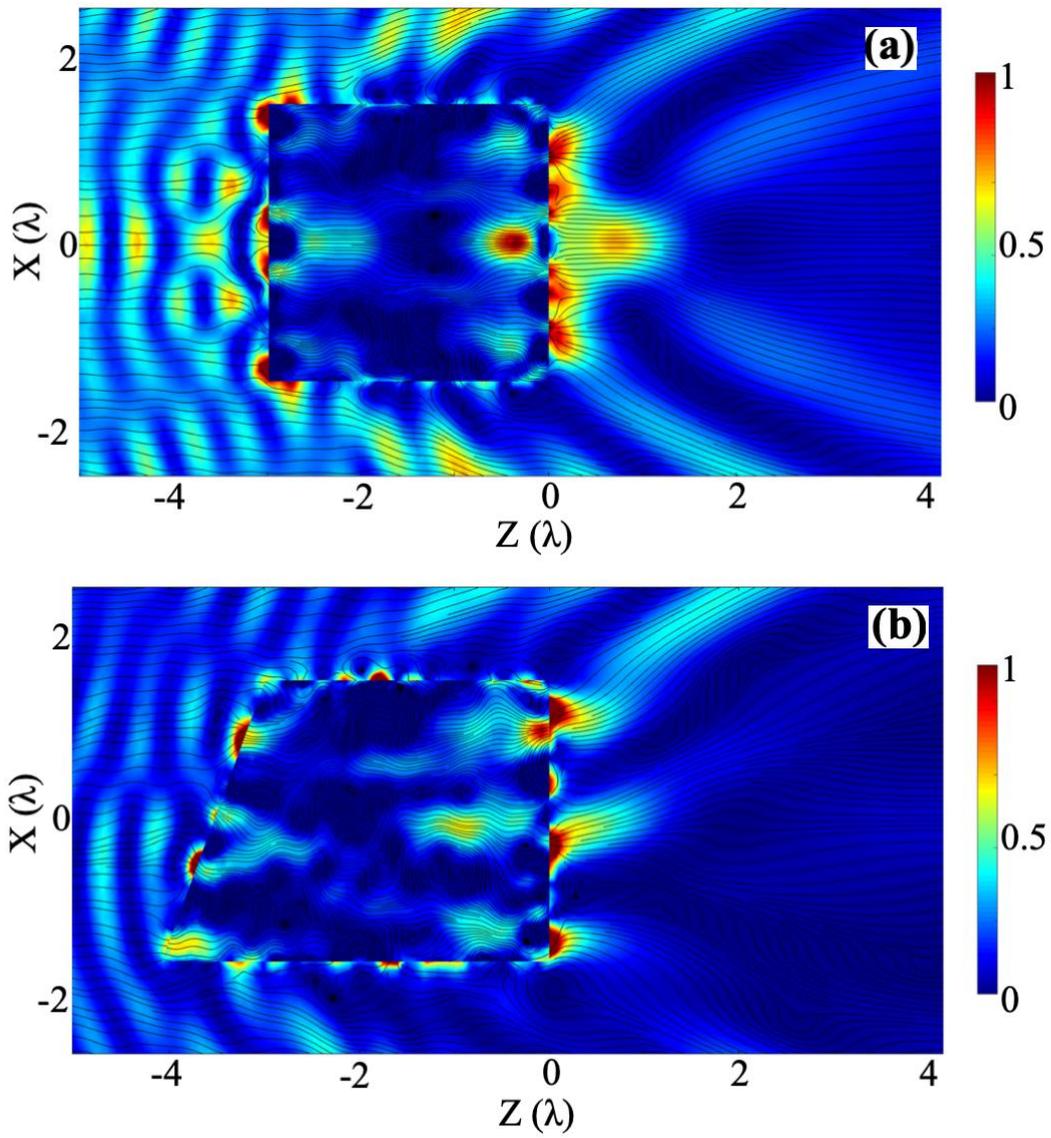

**FIGURE 5**